\documentclass[epj]{svjour}
\usepackage{amssymb,amsmath,latexsym,fancyhdr,bbold,epsfig,graphicx}
\usepackage{epic,eepic}
\usepackage[latin1]{inputenc}
\usepackage[squaren]{SIunits}

\renewcommand{\vec}[1]{\boldsymbol{#1}}
\newcommand{\mat}[1]{\boldsymbol{\sf #1}}

\renewcommand{\d}{\text{d}}
\renewcommand{\@}{\partial}
\newcommand{\e}{\text{e}}
\newcommand{\m}[1]{\left\langle #1 \right\rangle}
\begin{document}
\title{Force-extension curves of bacterial flagella}
\author{Reinhard Vogel \inst{1} \and Holger Stark\inst{1}
}                     
%
%
\institute{Institute for Theoretical Physics , TU Berlin \and Institute for Theoretical Physics , TU Berlin}
\date{Received: date / Revised version: date}
%
\abstract{
Bacterial flagella assume different helical shapes during the tumbling
phase of a bacterium but also in response to varying environmental 
conditions. Force-extension measurements by Darnton and Berg explicitly 
demonstrate a transformation from the coiled to the normal helical state 
[N.~C. Darnton and H.~C. Berg, Biophys. J.\ \textbf{92}, 2230 (2007)]. 
We here develop  an elastic model for the flagellum based on Kirchhoff's 
theory of an elastic rod that describes such a polymorphic transformation
and use resistive force theory to couple the flagellum to the aqueous 
environment. We present Brownian dynamics simulations that quantitatively
reproduce the force-extension curves and study how the ratio $\Gamma$
of torsional to bending rigidity and the extensional rate influence the 
response of the flagellum. An upper bound for $\Gamma$ is given. Using
clamped flagella, we show in an adiabatic approximation that the 
mean extension, where a local coiled-to-normal transition occurs first,
depends on the logarithm of the extensional rate.
\PACS{
      {PACS-key}{discribing text of that key}   \and
      {PACS-key}{discribing text of that key}
     } 
} 
\maketitle
\section{Introduction} \label{intro}

Many types of bacteria, such as \textit{Escherichia coli} and 
\textit{Sal\-mo\-nella typhimurium}, propel themselves forward
by rotating a bundle of elastic filaments with helical shape. Each
of these flagella, as they are called, is driven by a rotary motor. 
When the sense of rotation of one motor is reversed, the attached 
flagellum leaves the bundle and undergoes a sequence of different
helical configurations characterized by their pitch, radius and 
helicity. During the flagellar polymorphism, the bacterium tumbles
and then continues swimming in a different direction, when the flagellum 
assumes its original form and returns into the bundle. Whenever the
bacterium senses a positive food gradient, it prolongs the swimming 
phase and thereby increases the food uptake\ \cite{Berg04}.
In recent years, bacteria and their flagella have also been used
in studies relevant to microfluidics to transport colloids 
\cite{Zhang2009,Behkam2007} and pump fluids \cite{Kim2008}
but also to create new liquid-crystalline phases of 
screw like objects \cite{Barry2006}. 
An understanding of the bacterial tumbling motion and the applications
of bacterial flagella in nanotechnology and microfluidics requires
a sufficiently simple elastic model that includes the flagellar polymorphism.
This article aims to provide such a model.

The bacterial flagellum consists of three parts. The rotary motor is 
embedded in the cell membrane and transmits its motor torque to the long 
helical filament with the help of the short and very flexible proximal 
hook. The filament of \textit{E.coli} bacteria is up to 10 $\micro\meter$ long 
and is about 0.02 $\micro\meter$ in diameter. It is relatively stiff but 
can switch between distinct polymorphic forms. The filament 
assumes these forms in response to external perturbations such as
changes in pH value, salinity, and temperature of the surrounding fluid\
\cite{Kamiya76,Kamiya77,Hasegawa82}, the addition of alcohols
\cite{Hotani80} or sugars \cite{Seville93}, and by applying
external forces or torques to the filament 
\cite{Macnab77,Hotani82,Darnton07a,Darnton07}.

\begin{figure}
 \begin{center}
\includegraphics[width=0.85\columnwidth]{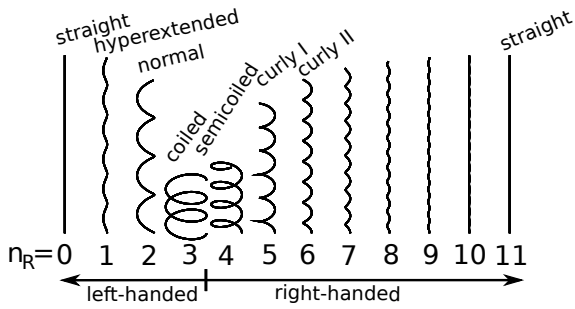}
\end{center}
\caption{Twelve polymorphic states of the bacterial flagellum with
curvature $\kappa=\kappa_\mathtt{max} \sin(\pi n_R/11)$ and torsion 
$\tau = \tau_L + (\tau_R-\tau_L) n_R/11$ after Calladine \cite{Calladine75},
where $n_R$ is the number of protofilaments in the R state.
The quantities $\kappa_\mathtt{max}\approx 2.4\per\micro\meter$, 
$\tau_L\approx-5.2\per\micro\meter$, and $\tau_R\approx11.8\per\micro\meter$ 
are fit parameters.}
\label{fig.polym}
\end{figure}

The helical filament is a cylinder formed by 11 protofilaments each of which
consists of a stack of protein mono\-mers called flagellin. These monomers 
assume two conformations (L and R) that mainly differ in length. Each 
protofilament only contains one type of monomer and therefore
L- and R-state protofilaments have different lengths.\footnote{This is one
  important ingredient of the model developed by Asakura and Calledine. The 
different helical configurations following from this model were confirmed 
later for example in Ref. \cite{Yamashita98}.}
Mixing them in the flagellar filament produces bending which together with an intrinsic 
twist of the filament gives a helical configuration. In total two straight 
and 10 helical polymorphic states are possible since protofilaments
of one type cluster (see Fig.\ \ref{fig.polym}). Most of them were 
observed experimentally\ \cite{Hasegawa1998}.
This is the picture Asakura \cite{Asakura70} and later 
Calladine\ \cite{Calladine75}, with a more detailed geometric model, 
developed to explain the flagellar polymorphism.
The molecular structure for the flagellar filament was confirmed recently
\cite{Yamashita98,Yonekura03} and several extensions of Calladine's model 
exist 
\cite{Hasegawa1998,Srigiriraju2005,Srigiriraju2006,Friedrich2006,Speier09}.

%

This article is mainly motivated by recent experiments of Darnton and 
Berg \cite{Darnton07}. With the help of an optical tweezer set up they 
pulled the two ends of the flagellar filament apart with a constant 
velocity and induced a transition between two polymorphic configurations.
They then reversed the velocity to compress the flagellum in order to
return it to the initial configuration. Darnton and Berg recorded 
force-extension curves mainly for the transformation from the coiled
to the normal configuration. The transformation starts locally at one end 
of the flagellum and then proceeds in discrete steps along the flagellum.
Signatures of the steps are clearly visible in the force-extension curves.


Calculating the force-extension curves on the basis of coarse-grained 
molecular dynamics simulations is not possible since simulation times 
are far below the experimentally relevant time scale of one second\ 
\cite{arkh:06}. Modeling the polymorphism of a bacterial flagellum on
a mesoscopic level is more appropriate. Goldstein \textit{et al.} 
extended Kirchhoff's classic theory of an elastic rod by introducing
a double-well potential for the spontaneous torsion to describe the 
transition between two helical states \cite{Goldstein2000,Coombs2002}.
Wada and Netz also described the helical filament by Kirchhoff's rod theory
but attached a spin variable along the filament in order to distinguish
locally between the two helical states\ \cite{Wada2008}. They then 
performed hybrid Brownian-dynamics Monte-Carlo simulations to numerically
calculate force-extension curves of bacterial flagella.


In this article we present Brownian-dynamics simulations of the 
force-extension curves
based on a model that is less time-con\-su\-ming than the approach of 
Ref.\ \cite{Wada2008} but that is completely equivalent as we demonstrate in
an appendix. We furthermore concentrate on different aspects of the
force-extension curves, namely how they depend on the ratio of torsional
and bending rigidities $\Gamma$ and on the velocity or extensional rate
with which the flagellum is pulled apart. We also give an upper bound 
for $\Gamma$ which is partially in contrast to experimental results. 
The mean extension, at which a coiled-to-normal transition first occurs 
locally, is a function of the extension rate. We demonstrate how this
extension can be inferred from equilibrium properties of a clamped 
helical filament. Our modeling is in the spirit of 
Refs.\ \cite{Goldstein2000,Coombs2002}. However, we show that a conventional
double-well potential cannot reproduce the experimentally observed
force extension curves. We therefore developed an alternative model, where
we just ``glue'' the harmonic elastic energies of the two helical states
together. We perform our simulations with realistic parameter values and 
can directly compare our results to experiments.

%

The article is organized as follows. In Section\ \ref{sec:1} we present
our extension of Kirchhoff's elastic rod theory to include the
polymorphism of helical filaments, explain how we perform the Brownian
dynamics simulations, and present analytic results for a uniformly 
stretched conventional helical filament. Section\ \ref{sec.forceext}
discusses the results for the simulated force-extension curves and
addresses the clamped filament. We close with a summary and conclusions
in Section\ \ref{sec.sum}.

%
%
%
%
%
%

\section{Continuum model for the bacterial flagellum} \label{sec:1}

\subsection{Classic theory of an elastic rod}
\begin{figure}
\begin{center}
\includegraphics[width=0.7\columnwidth]{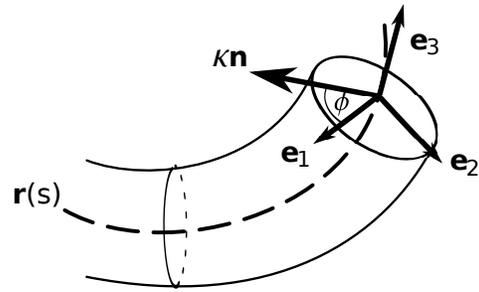}
\end{center}
\caption{The conformation of a slender elastic rod 
is described by the space curve $\vec r(s)$ of its center line
and the material frame $\{\vec e_1,\vec e_2,\vec e_3\}$. 
The vector $\kappa \vec{n} = \@_s\vec e_3$ describes the local 
curvature of $\vec r(s)$.}
\label{fig.material}
\end{figure}

The conformation of a slender rod with contour length $L$ is described by 
the space curve of its center line $\vec r(s)$, parametrized by the arc 
length $s$, and a material frame of three orthogonal unit vectors 
$\{\vec e_1,\vec e_2,\vec e_3\}$, attached to each point on the center line. 
The vector $\vec e_3$ points along the tangent of $\vec r(s)$ and 
$\vec e_1, \vec e_2$ typically correspond to the principal axes of the 
inertia tensor of the cross section as indicated in 
Fig.\ \ref{fig.material}.\footnote{For a circular cross section the 
eigenvalues of the inertia tensor degenerate and we are free to choose 
the vectors $\vec e_1$ and $\vec e_2$.}
Since the $\vec e_i$ are unit vectors,
the transport of the material frame along the center line is described by 
the generalized Frenet-Serret equations 
\begin{align}
\@_s \vec e_i=\vec \Omega \times \vec e_i ,
\end{align}
where $\@_s$ means derivative with respect to $s$. So the conformation of 
an elastic rod or filament is completely characterized by the angular strain vector $\vec \Omega = (\Omega_1,\Omega_2,\Omega_3)$.
Alternatively, one uses the Frenet frame consisting of the tangent vector
$\vec t =\vec e_3$, the curvature vector
$\kappa \vec{n} = \@_s \vec t $, and the binormal $\vec b = \vec t \times
\vec{n}$. The Frenet frame transforms into the material frame when it is
rotated about $\vec t =\vec e_3$ by the twist angle $\phi$, as indicated 
in Fig.\ \ref{fig.material}.
This relates the vector $\vec \Omega$ to the curvature $\kappa$ and 
torsion $\tau$ of the filament:
\begin{align}
\Omega_1&= \kappa \sin \phi, & \Omega_2&=\kappa \cos \phi, & \Omega_3&
= \tau +\@_s \phi ,
\label{eq.omega}
\end{align}
where the components $\Omega_i$ are given with respect to the local
material frame.


Kirchhoff's theory expands the elastic free energy $\mathcal F$ of a
deformed elastic rod up to second order in the angular strain $\vec \Omega$,
\begin{align} 
\label{Glg_freieEnergie1}
 {\mathcal F}= &\int_0^L f_{\mathtt{cl}}(\vec \Omega, \vec \Omega_0) 
\mathrm d s\\
\label{Glg_freieEnergie2}
f_{\mathtt{cl}}(\vec \Omega, \vec \Omega_0)&=\frac{A}{2} (\Omega_1)^2 
+ \frac{A}{2} (\Omega_2-\kappa_0)^2 +\frac{C}{2} (\Omega_3-\tau_0)^2 ,
\end{align}
where we also introduced a spontaneous curvature $\kappa_0$ and 
torsion $\tau_0$ within $\vec \Omega_0=(0,\kappa_0,\tau_0)$, and 
$A$ is the bending rigidity and $C$ the torsional rigidity
\cite{Love1944,Landau1986}. Since the bacterial flagellum has a circular
cross section, we can choose for the material frame the Frenet frame in 
the undeformed ground state, meaning $\phi =0$, and thereby restrict 
the spontaneous curvature $\kappa_0$ to $\Omega_2$. 
If $\kappa_0$ and $\tau_0$ are constant along the filament,
it has a helical shape with pitch $p=2 \pi \tau_0/(\kappa_0^2+\tau_0^2)$ and 
radius $r=\kappa_0/(\kappa_0^2+\tau_0^2)$. Note that although free energy 
(\ref{Glg_freieEnergie1}) is formulated in the spirit of linear elasticity 
theory, it becomes highly nonlinear when expressed in terms of the
space curve $\vec r(s)$ and twist angle $\phi$.


\subsection{Extended Kirchhoff rod theory}

In order to describe the transition of the bacterial flagellum between
two polymorphic configurations, the Kirchhoff rod theory has to be
extended to include two local ground states characterized by
$\vec \Omega_i=(0,\kappa_i,\tau_i)$ ($i=1,2$). We first tried to
generalize the approach of Goldstein \textit{et al.}, who used a typical
double well potential for the twist density $\Omega_3$ 
\cite{Goldstein2000,Coombs2002} (see Appendix\ \ref{sec.doublewell}) . 
However, the force-extension curves calculated with this extended free energy 
by numerical simulations did not reproduce the experimental curves as 
indicated in Appendix\ \ref{sec.doublewell}.

\begin{figure}[t]
  \begin{center}
        \includegraphics[width= \columnwidth]{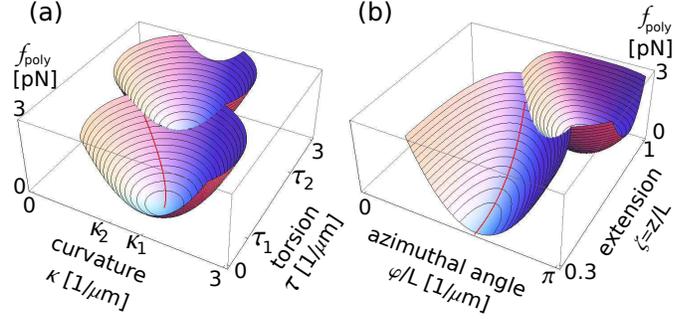}
  \end{center}
\caption{(a) Free energy density $f_\mathtt{poly}$ from Eq.\ (\ref{Glg Min})
as a function of curvature $\kappa$ and torsion $\tau$ using 
$\vec \Omega = (0,\kappa,\tau)$. (b) Free energy density $f_\mathtt{poly}$
as a function of height $z$ and azimuthal angle per unit length
$\varphi / L$ of a uniformly stretched helical filament 
(see Section\ \ref{subsec.homo}). The red lines indicate the path of the 
helical filament in the energy landscape during stretching.}
\label{fig.elastic}
\end{figure}

We, therefore, developed a different model. To each of the two relevant
polymorphic forms of the flagellum we assign the elastic free energy 
(\ref{Glg_freieEnergie1}) of Kirchoff's rod theory and introduce a 
difference $\delta$ of the two energy densities in the ground states. Since 
the free energy of a system always assumes a minimum, we locally assign 
to the bacterial flagellum with angular strain $\vec \Omega$
the minimum $f_\mathtt{poly}(\vec \Omega, \vec \Omega_1, \vec \Omega_2)$ 
of the free energy densities of the two polymorphic configurations:
\begin{align}\label{Glg Min}
f_\mathtt{poly}(\vec \Omega, \vec \Omega_1, \vec \Omega_2) = 
\min\big(f_\mathtt{cl}(\vec \Omega, \vec \Omega_1),
f_\mathtt{cl}(\vec \Omega, \vec \Omega_2)+\delta\big),
\end{align}
where $f_\mathtt{cl}$ is the elastic free energy density of 
Eq.\ \eqref{Glg_freieEnergie2}. The resulting density 
$f_\mathtt{poly}(\vec \Omega, \vec \Omega_1, \vec \Omega_2)$ is plotted in 
Fig.\ \ref{fig.elastic} as a function of curvature $\kappa$ and torsion 
$\tau$ using Eq.\ (\ref{eq.omega}) and assuming $\phi=0$.

Wada and Netz formulated an alternative, statistical model to access the
polymorphism of the bacterial flagellum \cite{Wada2008} which they described 
as a bead-spring chain. To each bead they assigned the Kirchoff free energy 
(\ref{Glg_freieEnergie2}) and an Ising spin to distinguish locally 
between the two polymorphic forms of the flagellum. The Ising spin 
Hamiltonian then favored the same polymorphic state for adjacent
beads. By integrating out the spin degree of freedom, they derived an effective 
free energy density which we summarize in Appendix\ \ref{App_Energy}. 
In experiments, where a thermally induced transition from the 
\textit{normal} to the \textit{semi-coiled} configuration was studied, 
most flagella assumed a pure polymorphic form of either the normal or 
semi-coiled state \cite{Hasegawa82} suggesting that the energy cost 
for forming a domain wall between the two helical states is much larger 
than thermal energy. Using this observation, we demonstrate in 
Appendix\ \ref{App_Energy} that the free energy density of Wada and 
Netz simplifies to our elastic free energy \eqref{Glg Min}.

%
%

Note that our ansatz can easily be extended to describe more than two of 
the polymorphic states of bacterial flagella. In particular, this will be 
necessary for understanding the sequence of polymorphic transitions 
during the tumbling phase of \textit{E. coli}.


The elastic free energy density $f_\mathtt{poly}$ of Eq.\ (\ref{Glg Min})
admits that two domains of different polymorphic states are separated
from each other by a sharp domain wall with zero width. To realize a
more realistic, smoother transition between two domains, we have to 
introduce a free energy density of the form \cite{Goldstein2000}
\begin{align}
 f_\mathtt{bi}(\@_s\vec \Omega)=\frac{\gamma}{2} \left(\@_s \vec \Omega\right)^2,
\label{eq.domain}
\end{align}
where $\gamma^{1/2}$ is proportional to the width of the domain wall.
With a properly adjusted $\gamma$, experimental observations of the 
domain wall could be described quantitatively \cite{Goldstein2000,Hotani82}. 
Furthermore, our simulations demonstrated that the helical 
filament contains several domains instead of just two when $\gamma$
is chosen too small or even equal to zero.

Instead of implementing a constraint for the filament to be 
inextensible, we introduce a stretching free energy density 
$f_\mathtt{st}= K \left(\@_s \vec r\right)^2/2$. We choose the spring
constant $K$ such that the changes in the filament length are below
1.5 \%. So the filament is inextensible to a good approximation.

Collecting all the contributions, the total elastic free energy
${\mathcal F} = \int_0^L f \mathrm ds$, which we will use in the 
following for modeling the bacterial flagellum, is based on the density
\begin{align}\label{Glg free energy}
 f=& f_\mathtt{poly}(\vec \Omega, \vec \Omega_1, \vec \Omega_2)
+f_\mathtt{bi}(\@_s\vec \Omega) +f_\mathtt{st}(\@_s\vec r).
\end{align}
Since $f$ ultimately depends on the centerline $\vec r(s)$ and the 
twist angle $\phi(s)$, the total free energy is a functional 
${\mathcal F}[\vec r(s),\phi(s)]$ in $\vec r(s)$ and $\phi(s)$.

%
%

\subsection{Dynamics of a Helical Rod}

We formulate Langevin equations for the location $\vec r(s)$ and 
intrinsic twist $\phi(s)$ of the helical filament.
At low Reynolds number the sum of elastic force per unit length,
$\vec f_\mathtt{el} = - \delta \cal F/\delta \vec r$, and thermal
force $\vec f_\mathtt{th}$ is balanced by viscous drag. The same
applies to the elastic torque per unit length, $m_\mathtt{el}=-\delta 
\cal F/\delta \phi$ and thermal torque $m_\mathtt{th}$.
Using resistive force theory \cite{Childress1981} that employs local 
friction coefficients $\gamma_\parallel, \gamma_\perp$ and $\gamma_R$
(see Appendix\ \ref{sec.friction}), we formulate the Langevin equations
\begin{align}
 \left[\gamma_\parallel \vec t \otimes 
\vec t + \gamma_\perp(\mat 1- \vec t \otimes \vec t )\right]
\vec v=&\vec f_\mathtt{el}+ \vec f_\mathtt{th}
\label{eq.dynamic1}\\
\gamma_R \omega = & m_\mathtt{el} + m_\mathtt{th} .
\end{align}
Here $\vec v=\@_t \vec r$ 
is the translational velocity, $\omega = \@_t \phi$ the angular 
velocity about the local tangent vector $\vec t=\vec e_3$, and 
$\otimes$ means tensorial product. The anisotropic friction tensor
acting on $\vec v$ in Eq.\ (\ref{eq.dynamic1}) couples rotation about the
helical axis to translation and thereby creates the thrust force that
pushes the bacterium forward \cite{Purcell1977}. The friction coefficients
per unit length were calculated by Lighthill from slender-body theory
taking into account the helical geometry of the rod \cite{Lighthill1976}. 
The coefficients are summarized in Appendix\ \ref{sec.friction}.
In experiments one finds reasonable agreement with this approach
\cite{Chattopadhyay2006,Chattopadhyay2009}.
The thermal force $\vec f_\mathtt{th}$ and torque $m_\mathtt{th}$ are
Gaussian stochastic variables with zero mean,
$\langle \vec f_\mathtt{th} \rangle = \vec 0$ and 
$\langle m_\mathtt{th} \rangle = 0$. Their variances obey the 
fluctuation-dissipation theorem and therefore read
\begin{align}
 \m{\vec f_\mathtt{th}(t,s) \otimes \vec f_\mathtt{th}(t',s')} &= 2 k_B T 
\delta(t-t')\delta(s-s') \\
 & \times \left[\gamma_\parallel^{-1} \vec t \otimes \vec t + \gamma_\perp^{-1}
(\mat 1- \vec t \otimes \vec t )\right],
\nonumber \\
\m{m_\mathtt{th}(t,s)m_\mathtt{th}(t',s')}&= 2 k_B T \delta(t-t')
\delta(s-s')\gamma_R^{-1},\\
\m{m_\mathtt{th}(t,s)\vec f_\mathtt{th}(t',s')}&=\vec 0.
\end{align}


\subsection{Details of Simulations}

In our simulations we used a technique developed by Rei\-chert 
\cite{Reichert2006} similar to the one of Chirico and Langowski 
\cite{Chirico1994} and also employed by Wada and Netz in simulations
of helical nano springs etc. \cite{Wada2007a,Wada2007}. We discretize
the centerline $\vec r(s)$ of the filament by introducing $N+1$ beads 
at locations $\vec r^i=\vec r(s=i\cdot h)$ and with nearest-neighbor 
distance $h$. To every bead we attach the material frame, i.e.,
a right-handed tripod of orthonormal vectors 
$\{\vec e^i_1,\vec e^i_2,\vec e^i_3\}$
($i=0,\dots,N$), where the tangent vector is approximated by
\begin{align}
 \vec e_i^3 &=\frac{\vec r_i-\vec r_{i-1}}{|\vec r_i-\vec r_{i-1}|}.
\end{align}
We split the rotation of the tripod along the filament into two parts. 
First, we rotate about the bond direction $\vec e_i^3$ by an angle 
$\Omega_i^3 h$ corresponding to the intrinsic twist plus torsion. Thereafter, 
the tripod is rotated such that the bond orientation $\vec e_3^{i}$ is 
transformed into the consecutive direction $\vec e_3^{i+1}$, thus 
describing the curvature of the filament. 
With this procedure the free energy density $f$ of 
Eq.\ (\ref{Glg free energy}) is discretized
and the functional derivatives of the total free energy,
$\vec f_\mathtt{el} = - \delta \cal F/\delta \vec r$ and 
$m_\mathtt{el}=-\delta \cal F/\delta \phi$, reduce to conventional 
derivatives with respect to $\vec r^i$ and $\phi^i$. We also note
that the tangent vector in the friction tensor in Eq.\ (\ref{eq.dynamic1})
is approximated by 
$\vec t^i = (\vec e_3^i + \vec e_3^{i+1})/|\vec e_3^i + \vec e_3^{i+1}|$. 

%

In the following, we will discuss the influence of the three relevant
parameters on the force-extension curves; the ratio $\Gamma=C/A$ of the
torsional and bending rigidities (twist-to-bend ratio $\Gamma$), 
the difference in energy $\delta$ of 
the two helical ground states under study, and the velocity or extensional
rate $v_p$ with which the bacterial flagellum is pulled apart. 
The bending rigidity $A$ 
together with an appropriate length introduces characteristic values 
for force and elastic energy. We used $A=3.5 \pico\newton \micro\meter^2$ 
given in Ref. \cite{Darnton07} as a typical value for bacterial 
flagella. All other parameters are determined by the geometry of the 
two polymorphic states. Initially, the bacterial flagellum is in the coiled 
state with spontaneous curvature $\kappa_1=1.8\per\micro\meter$ and torsion 
$\tau_1=0.56\per\micro\meter$ and then switches into the normal state 
with $\kappa_2=1.3\per\micro\meter$ and $\tau_2=2.1\per\micro\meter$.

The friction coefficients are calculated with the formulas of Lighthill 
\cite{Lighthill1976} summarized in Appendix\ \ref{sec.friction} as 
$\gamma_\parallel=1.6\cdot 10^{-3} 
\pico\newton\second\per\micro\meter^2$, $\gamma_\perp=2.8\cdot10^{-3} 
\pico\newton\second\per\micro\meter^2$, and $\gamma_R=0.126\cdot 
10^{-3} \pico\newton\second$ where a filament diameter of around 
$20\nano\meter$ was used. The length of the filament is $L=10\micro\meter$ 
corresponding to approximately three helical turns in the coiled and four 
in the normal state. The discretization length between the beads was 
chosen as $h=0.2\micro\meter$ and the elastic coefficient in 
Eq.\ (\ref{eq.domain}) as $\gamma=0.1 \pico\newton \micro\meter^4$.

%
%

\subsection{Uniform Deformation} \label{subsec.homo}

A helical filament of length $L$ much larger than the pitch deforms 
uniformly aside from inhomogeneities at both ends when a constant 
external force pulls at it.\footnote{For a force compressing the helical filament one observes buckling \cite{Goriely1997}.}
This means curvature $\kappa$ and torsion $\tau$ are 
constant along the filament. We orient the helical axis along the
$z$ axis of a cylindrical coordinate system $(\rho,\varphi,z)$. The
geometry and free energy of the uniformly deformed helix are completely 
described by its height $z= L \frac{\tau}{\sqrt{\kappa^2+\tau^2}}$ and 
the difference in azimuthal angles of both ends of the filament, 
$\varphi=\varphi(s=L) = L \sqrt{\kappa^2+\tau^2}$, 
where we set $\varphi(s=0)=0$. Then, the force and torque acting on 
a uniformly stretched helix follows from
$F=-\partial_z \mathcal F(z,\varphi)$ and 
$T=-\partial_\varphi \mathcal F(z,\varphi)$, respectively.
 In experiments,
the ends of the filament can freely rotate during stretching which means 
zero torque $T$. 
This leads to an expression for $\varphi/L$
\begin{align}\label{Glg_phiL}
 \frac{\varphi}{L}&=\frac{\kappa_0\cos\alpha +\Gamma\tau_0\sin\alpha}{(1+\Gamma) \cos^2\alpha}
\end{align}
where we introduced the pitch angle $\alpha$ using the extension 
\begin{equation}
\zeta= z/L = \sin \alpha .
\end{equation}

Under the condition of zero torque, we then obtain the classic force-extension 
relation \cite{Love1944}
\begin{align}
\label{Glg homogen}
  F(z) &=\frac{A(\Gamma \kappa_0 \cos\alpha  +\tau_0\sin\alpha)
(\tau_0\cos\alpha-\kappa_0 \sin\alpha)}{\cos\alpha( \Gamma \cos^2\alpha  
+\sin^2\alpha)^2} \\
 &\approx A \frac{ (\kappa_0^2 + \tau_0^2)^3 }{\kappa_0^2 ( \Gamma \kappa_0^2 
+ \tau_0^2)}\frac{z_0-z}{L}:=- k \frac{z-z_0}{L}\nonumber ,
\end{align}

In the second line of Eq.\ (\ref{Glg homogen}), the force is linearized 
in a small relative extension $(z-z_0)/L$, where $z_0$ is the height of 
the undeformed helix and $k$ the spring constant. 
Interestingly, we find that curvature 
$\kappa(\zeta)=\sqrt{1-\zeta^2} \varphi/L$ and 
torsion $\tau(\zeta)=\zeta \varphi/L$ of a helix with
extension $\zeta=z/L$ lie on the ellipse
\begin{align} 
\label{Glg_Ellipse}
[\kappa(\zeta)-\kappa_0/2]^2 + \Gamma[\tau(\zeta)-\tau_0/2]^2 & = 
(\kappa_0/2)^2 + \Gamma(\tau_0/2)^2,
\end{align}
the center of which is at $(\kappa_0/2,\tau_0/2)$. In deriving Eq. \eqref{Glg_Ellipse} we used expression \eqref{Glg_phiL} for $\varphi/L$. For a symmetric 
potential with twist-to-bend ratio $\Gamma=1$, the ellipse becomes
a circle. We use relation (\ref{Glg_Ellipse}) in Figs.\ \ref{fig.elastic}(a), 
\ref{Fig_A/C_Vergleich}(b), and \ref{fig.extendedall} to indicate the path 
of a uniformly stretched flagellum in the $\kappa,\tau$ plane.

Bacterial flagella only have a length of a few pitches. So finite size
effects from both ends lead to a noticeable dependence of curvature and 
torsion on arc length $s$. Nevertheless, we find that
Eq.\ (\ref{Glg homogen}) is a good approximation for the initial part
of the force-extension curve. This is valid since the experimental
pulling rate is so small that the extending filament passes through a
sequence of equilibrium configurations as we demonstrate in the following
section.

\subsection{Characteristic Time Scale and Velocity} 
\label{subsec: TimeScale}

Applying instantaneously a force at one end of the helical filament 
induces a localized deformation which spreads along the helix. 
To treat this situation approximately, we assume the helical filament
to be an extensible rod that locally obeys the linearized force-extension 
relation of Eq.\ \eqref{Glg homogen} with $(z-z_0)/L$ replaced by $\d z/\d s$.
In addition, the rod possesses the local friction coefficient per unit 
length for moving a helix parallel to its axis, $\gamma_H = 
\gamma_\perp - (\gamma_\perp-\gamma_\parallel) (\d z/\d s)^2$.
Balancing elastic and viscous forces locally, gives the diffusion equation
\begin{align}
 \@_t z &= \frac{k}{\gamma_H} \@_s^2 z,
\end{align}
where $k$ is the spring constant of the helix given in Eq.\ 
\eqref{Glg homogen}. A localized deformation at one end of the helical 
filament spreads diffusively over the whole filament on a time scale 
$\tau_C=\gamma_H L^2/k \approx 10^{-4}\second$ which gives the 
characteristic velocity
$v_C = L/\tau_C \approx 10^3 \micro \meter \per \second$. In experiments
but also in our simulations one pulls at the bacterial flagellum with
velocities much smaller than $v_C$. So the extending filament passes 
through a sequence of equilibrium configurations. This also means that 
the applied and elastic forces nearly balance each other since they
are much larger than the frictional forces.
Note that a diffusion equation that balances elastic and viscous 
forces shows up as well in the twist dynamics of a rotating elastic 
filament \cite{Wolgemuth2000}.




\section{Force-extension curves} \label{sec.forceext}

\subsection{Pulling on and compressing the helical filament}

We performed both Stokesian (temperature $T=0$) and Brownian dynamics
($T=300K$) simulations  of the force-extension measurements in
Ref.\ \cite{Darnton07}. Similar to the experimental setup, we fix one 
end of the filament whereas the other end is allowed to move in a
harmonic potential with trap stiffness $100 \,\mathrm{pN/\mu m}$
(as in Ref.\ \cite{Darnton07}) mimicking the experimental situation 
where a bead 
attached to the bacterial flagellum was trapped by an optical tweezer.
The axis of the helical filament is oriented parallel to the $z$ axis.
In the beginning, the minimum of the harmonic potential coincides with 
one end of the filament in the initial coiled state. We then move the 
potential with a constant velocity $v_p$
along the $z$ axis and the filament stretches. After reaching a 
maximum extension $\zeta_M=z_M/L$,
the potential moves with the opposite velocity $-v_p$ back into the 
initial position. Note that in experiments the extension rates were
$v_p=0.4 \micro\meter\per\second$ and less \cite{Darnton07}. Such
small velocities are very time consuming in our simulations so that
we typically chose values from the range  
$v_p=2,\dots,20\micro\meter\per\second$ for recording the complete
force-extension curve. When we were just monitoring the initial part of 
the curve, we used velocities as small as $v_p=0.2 \micro\meter\per\second$ 
(see Section\ \ref{subsec.clamped}).

%



\begin{figure}
\begin{center}
\includegraphics[width=1.\columnwidth]{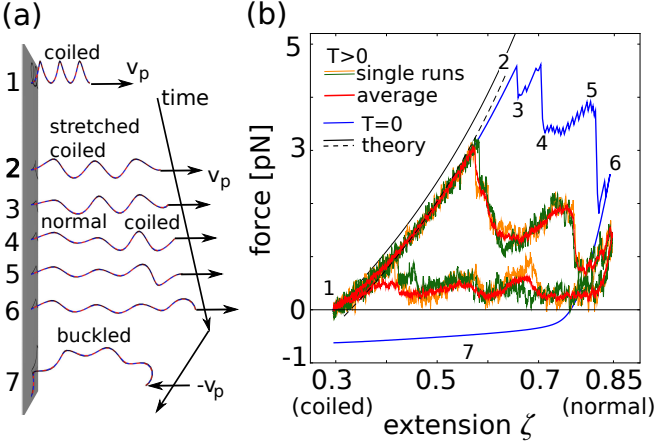}
\end{center}
%
%
\caption{(a) Snapshots of a helical filament stretched at $T=0$
with velocity $v_p = 2 \mu \mathrm{m}/\mathrm{s}$ taken from the 
supplementary movie 1. The coiled-to-normal
transition ($v_p>0$) and buckling ($v_p<0$) are visible. 
(b) Force-extension curves simulated without thermal noise ($T=0$) 
(blue curve) and with thermal noise: two realizations 
(thin orange and green lines) and an average over 10 runs (thick red line)
are shown.
The parameters are $v_p=2 \mu \mathrm{m}/\mathrm{s}$ and $\Gamma=0.7$.
Analytic prediction for a uniformly stretched filament (thin black line)
and shifted curve (dashed black line).}
\label{fig_fe1}
\end{figure}

We first simulated the extension of a filament with only one helical
state. Similar to the snapshots 1 and 2 in Fig.\ \ref{fig_fe1}(a) the
deformation of the filament is uniform except for small regions at
both ends. This leads to a small deviation of the simulated force-extension 
curve from the theoretical prediction of Eq. (\ref{Glg homogen})
for a uniformly stretched filament. The
situation is similar to the results in Fig.\ \ref{fig_fe1}(b), where 
the initial part 1-2 of the simulated blue curve is compared to the 
analytic prediction (thin black line). If we even shift the thin black 
line to the right, the dashed black line agrees very well with the blue 
curve besides the initial part close to position 1. We checked that 
even this difference gradually vanishes with increasing height of the 
helix as expected. In Section\ \ref{subsec: TimeScale} we already 
explained that our simulations are performed in the quasistationary
regime. This is also confirmed by our observation that at an extension
rate $v_p=2\micro\meter\per\second$ the dissipated energy during one 
extension-and-compression cycle is only about $2\%$ of the maximum 
elastic energy at extension $\zeta_M$. Furthermore, we do not 
observe any pronounced difference between deterministic Stokesian 
and Brownian dynamics simulations. For $T \ne 0$, the forces fluctuate
around a value which agrees with the deterministic force at $T=0$.


We then determined the force-extension curve at $T=0$ when the helical 
filament can switch from the coiled to the normal state using the 
elastic free energy\ \eqref{Glg free energy}. The results are illustrated 
in Fig.\ \ref{fig_fe1}, where the blue curve in (b) corresponds to $T=0$.
At a certain extension (position 2) the measured force drops sharply
since a small segment of the filament close to the fixed end switches
into the normal state as snapshots 2 and 3 in Fig.\ \ref{fig_fe1}(a)
reveal. Then the filament is stretched again. Further adjacent segments 
transform suddenly until at position 6 the filament is nearly completely 
in the normal state. From here we invert the velocity $v_p$ and move both ends 
together. However, the filament does not transform back into the coiled 
state but remains in the normal state. This ultimately leads to a 
negative force under which the filament buckles [see snapshot 7
in Fig.\ \ref{fig_fe1}(a)]. The full cycle is shown in the supplementary 
movie 1. 

Brownian dynamics simulations reveal the influence of thermal 
fluctuations on the force-extension curve. Two realizations
are included in Fig.\ \ref{fig_fe1}(b) as thin lines and the thick
red line shows an ensemble average over 10 runs. A full cycle of one 
realization is shown in the supplementary movie 2. Clearly, the
first transition into the normal state occurs at a smaller extension
compared to the deterministic case since thermal fluctuations help 
to overcome the energy barrier in the elastic free energy density
(see Fig.\ \ref{fig.elastic}). Whereas the force in each realization
fluctuates visibly, the sharp decrease of the force when a local transition
into the normal state occurs is as pronounced as in the deterministic case.
Finally, when both ends of the filament are moved together, it
completely transforms back to the coiled state and buckling is not
observed. The single realizations of the complete cycle of the
force-extension curve closely resemble the experimental curves in
Figs. 5 and 6 of Ref. \cite{Darnton07}. In particular, the force and 
extension where the first coiled-to-normal transition occurs fall into the
experimental ranges of 3 to 5 pN and $\zeta = 0.55$ to 0.6, respectively.

We now discuss in more detail how the difference $\delta$ in the 
ground-state energies of the two helical states, the twist-bend ratio $\Gamma$,
and the extension rate $v_p$ influence the force-extension curves.


\subsubsection{Ground-state energy difference $\delta$ of the coiled 
and normal state}

Increasing the ground-state energy $\delta >0$ of the normal relative to 
the coiled state also increases the energy barrier, which the flagellum 
in the coiled configuration has to overcome to transform locally into the 
normal state. Therefore, the transition is delayed to a larger extension $z/L$
or does not occur at all. On the other hand, the barrier which the normal 
configuration has to overcome to relax back into the coiled state
decreases and buckling of the filament becomes less probable.
Observations also show that a filament prepared in the normal state 
very slowly relaxes back into the coiled state, so $\delta$ should not be
too large compared to thermal energy $k_BT$. Using also the following 
quantitative considerations, we 
adjusted $\delta$ to $0.1 \pico\newton$ which resulted in the good 
agreement with experimental observations, already demonstrated.



We now derive an upper bound for $\delta$ to ensure that a 
transformation from the coiled to the normal state is, in principal, 
observable. The locus of the energy barrier in the $\kappa,\tau$ 
plane of Fig.\ \ref{fig.elastic} is given by 
\begin{align}
f_{cl}(\vec\Omega,\vec\Omega_1) =  f_{cl}(\vec\Omega,\vec\Omega_2) +\delta,
\end{align}
where $\vec\Omega=(0,\kappa,\tau)$ and $\vec\Omega_i = (0,\kappa_i,\tau_i)$
contains the values for spontaneous bend $\kappa_i$ and torsion $\tau_i$
of the coiled ($i=1$) and normal state ($i=2$), respectively. Assuming 
both states have the same elastic constants $A$, $C$, we arrive at a 
straight line
\begin{align} 
   \kappa  (\kappa_1-\kappa_2)+\tau \Gamma(\tau_1-\tau_2)+\delta/A & 
\nonumber\\
& \hspace*{-2.5cm} = (\kappa_1^2- \kappa_2^2)/2
+ \Gamma (\tau_1^2- \tau_2^2)/2 .
\label{Glg_Grenze}
\end{align}
We will use this formula later. The ground-state energy density of the 
normal state at $\vec\Omega_2 = (0,\kappa_2,\tau_2)$ is $\delta$. Only 
if $\delta$ lies below the energy density of the coiled state at 
$\vec\Omega = \vec\Omega_2$ is a clear transition between both states 
possible. The explicit form of this upper bound, 
$\delta < f_{cl}(\vec\Omega_2,\vec\Omega_1)$, combines with $\delta > 0$ 
to the inequality
\begin{align} 
\label{Glg_delta1}
0 < \frac{\delta}{A (\kappa_1-\kappa_2)^2} 
  < & (1+\Gamma \Delta^2) / 2 ,
\end{align}
where
\begin{equation} 
\label{Glg_delta2}
\Delta = \frac{\tau_1-\tau_2}{\kappa_1-\kappa_2}
\end{equation}
is the ratio of the differences in spontaneous torsion and curvature.

In experiments the value of $\delta$ changes with the conditions of the
solvent. In particular, different polymorphic forms of the bacterial
flagellum become stable when one alters the pH value, ionic strength, or 
temperature of the aqueous environment \cite{Hasegawa82}. It would be
interesting to perform the force-extension experiments under different
conditions to investigate how a changing $\delta$ but also variations
in bending ($A$) and torsional ($C$) rigidity influence the force-extension
curve of a bacterial flagellum.


\subsubsection{Twist-to-bend ratio $\Gamma$}


\begin{figure*}[t]
\begin{center}
\includegraphics[width=1.9 \columnwidth]{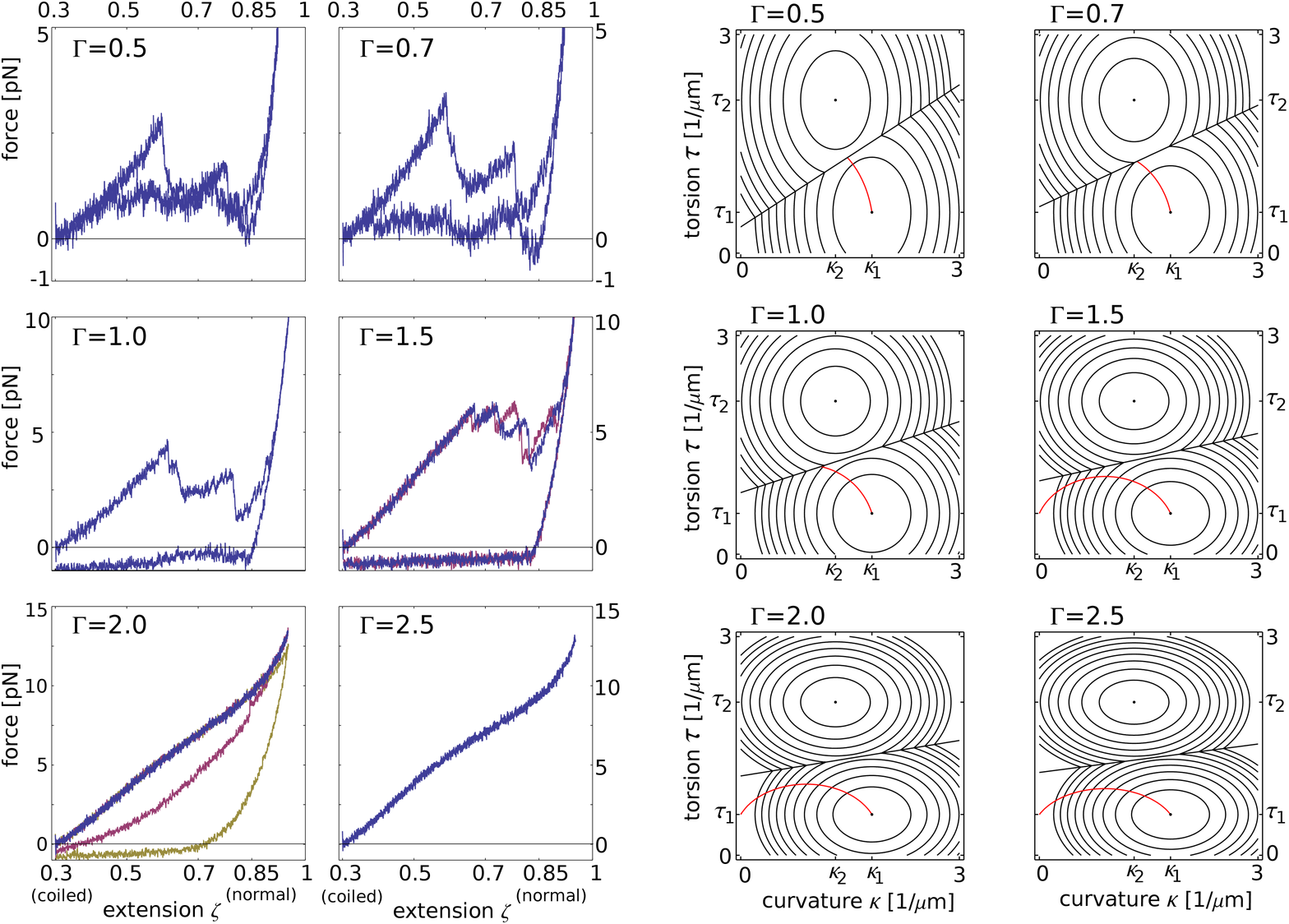}
\end{center}
\caption{(a) One or several realizations of force-extension curves
for increasing twist-to-bend ratios $\Gamma$. The extensional rate is 
$v_p = 2 \mu\mathrm{m}/\mathrm{s}$. (b) Contour plots of the elastic free 
energy density as a function of curvature and torsion for the same 
$\Gamma$ as in (a). The red line indicates the sequence of $\kappa, \tau$
values, as the filament in the coiled state is uniformly stretched.}
\label{Fig_A/C_Vergleich}
\end{figure*}

Increasing the twist-to-bend ratio $\Gamma$ also increases the energy 
barrier as the following formula for the minimum value of the barrier 
demonstrates,
 \begin{align}
\frac{f_b}{A(\kappa_1-\kappa_2)^2} = &
\frac{\left(1 
+ 2 \delta/[A(\kappa_1-\kappa_2)^2]
+\Gamma \Delta^2 \right)^2 }
{8  \left( 1 + \Gamma \Delta^2 \right)}.
 \end{align}
However, in contrast to $\delta$ an increasing $\Gamma$ increases both 
barriers for the coiled-to-normal and the normal-to-coiled transition.
We now discuss in 
Fig.\ \ref{Fig_A/C_Vergleich}(a)
how the twist-to-bend 
ratio $\Gamma$ influences the force-extension curve of the helical
filament. The filament is stretched to an extension $\zeta=0.95$, well
above the equilibrium height $\zeta=0.85$ of the normal state.
Fig.\ \ref{Fig_A/C_Vergleich}(b)
shows contour plots of the elastic free energy
density as a function of curvature and torsion for the same $\Gamma$ as in
(a). The red line indicates the sequence of $\kappa, \tau$
values, as the filament in the coiled state is uniformly stretched
(see Section\ \ref{subsec.homo}).
Note that for $\Gamma \ne 1$ the anisotropy of the elastic free energy 
density is clearly visible.



For $\Gamma=0.5$ and $0.7$, one observes the typical force-extension
curves as already discussed in Fig.\ \ref{fig_fe1} also for $\Gamma=0.7$. 
Since now the maximal extension $\zeta=0.95$ is larger, the curve for 
$\Gamma=0.7$ during compression looks different. While at $\zeta=0.95$ 
the entire filament is in the normal state, the filament for
$\Gamma = 1$ no longer transforms completely into the normal state.
Most pronounced at $\Gamma = 1$ is the fact that 
even with thermal forces the filament does not return 
into the coiled state during compression. This results in a negative 
force under which the filament buckles. All the force-extension curves in 
Fig.\ \ref{Fig_A/C_Vergleich}(a)
are determined for an extension rate
$v_p = 2\micro\meter\per\second$. A smaller $v_p$ enhances the probability
that thermal fluctuations transform the filament back into the coiled state
and buckling is not observed. The same is true for a smaller extension 
where a larger part of the filament stays in the coiled state and makes 
it easier for the rest of the filament to return to the
coiled state. At a ratio $\Gamma=1.5$ the force-extension curve changes 
qualitatively. A first transition into the normal state occurs at a larger
extension $\zeta\approx0.65$ compared to the previous curves. The 
transitions are no longer as pronounced and there are larger differences
between single realizations of the force-extension curve.
In addition, for $\Gamma < 1.5$ only one normal domain occurs 
whereas for $\Gamma=1.5$ two normal domains are observed. The reason for 
all these features becomes clear from Fig.\ \ref{Fig_A/C_Vergleich}(b).
At $\Gamma=1.5$ a uniformly stretched
filament does not hit the energy barrier anymore but passes the barrier 
in a close
distance so that thermal fluctuations are needed 
to induce transformations into the normal state. 
For further increase of
$\Gamma$ a transition into the normal state does not occur at all 
realizations, as the three graphs for $\Gamma=2$ demonstrate. A similar
behavior is observed in Ref. \cite{Darnton07} for a transition from
the normal into the hyperextended state.
The blue curve corresponds to the traditional force-extension relation of
the helix in the coiled state. In the yellow curve a complete transition
into the normal state was realized. A partial transformation occurred in the
magenta curve which then completely returned into the coiled state.
Finally, at $\Gamma=2.5$ the filament always remains in the coiled state.
Beyond an extension of $\zeta = 0.5$ the slope of the curve becomes smaller.
This is the onset of a qualitatively new behavior. At even larger $\Gamma$ 
and $\tau_0/\kappa_0 < 1$, one observes a sharp drop in the force due
to a discontinuous transformation, where one turn of the helical filament 
unwinds. This is discussed in Refs.\ \cite{Kessler2003,Wada2007a}.

\begin{figure}
\begin{center}
 \includegraphics[width=0.6\columnwidth]{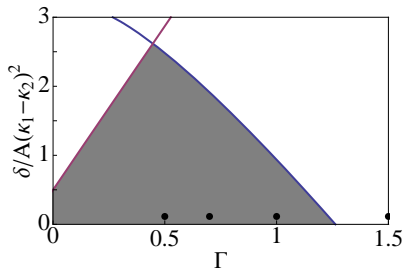}
\end{center}
\caption{The shaded area indicates the parameter ranges for
the ground-state energy difference $\delta$ and the twist-to-bend 
ratio $\Gamma$ for which a coiled-to-normal transition should be observed.
The dots give parameter values used in simulations. Note that thermal
fluctuations shift the upper border for $\Gamma$ indicated by the blue 
line to the right.}
\label{Fig_Gamma-Delta}
\end{figure}

As discussed, neglecting thermal fluctuations, a coiled-to-normal transition 
is no longer observable in the force-extension curve when in 
Fig.\ \ref{Fig_A/C_Vergleich}(b)
the straight line [Eq.\ (\ref{Glg_Grenze})] separating the coiled and 
normal state becomes tangential to the trajectory [Eq.\ (\ref{Glg_Ellipse})] 
of the uniformly stretched filament. Combining both equations leads to 
a second condition for $\delta/A$:
\begin{align}
\frac{\delta}{A (\kappa_1-\kappa_2)^2} & < \frac{1}{2 (\kappa_1-\kappa_2)} 
\bigg(\kappa_2 + \Gamma  \tau_2 \Delta  \nonumber\\
& \qquad \qquad + \sqrt{(1 + \Gamma \Delta^2) 
(\kappa_1^2+\Gamma  \tau_1^2)} \bigg)
\label{Glg_delta3}
\end{align}
Together with condition \eqref{Glg_delta1}, we obtain a region in the
parameter space $(\Gamma,\delta/A)$ where a transformation from the
coiled to the normal state should occur. The region is indicated as 
shaded area in Fig.\ \ref{Fig_Gamma-Delta}. Based on experimental data
for Young's and shear modulus in literature, Flynn and Ma received
for the twist-to-bend ratio $\Gamma$ the range 
$2 \cdot 10^{-1} - 2\cdot 10^2$ \cite{Flynn2004}. 
With a computational method, called
the quantized elastic deformational model, they calculated 
$\Gamma \approx 23$ \cite{Flynn2004}. On the other hand, together with 
our value $ \delta/[A (\kappa_1-\kappa_2)^2]\approx 0.11$, we predict a 
value of $\Gamma \lesssim 1$, in good agreement with our simulations.

\subsubsection{Extensional rate}

\begin{figure}[t] 
    \begin{center}
	\includegraphics[width=.49 \columnwidth]{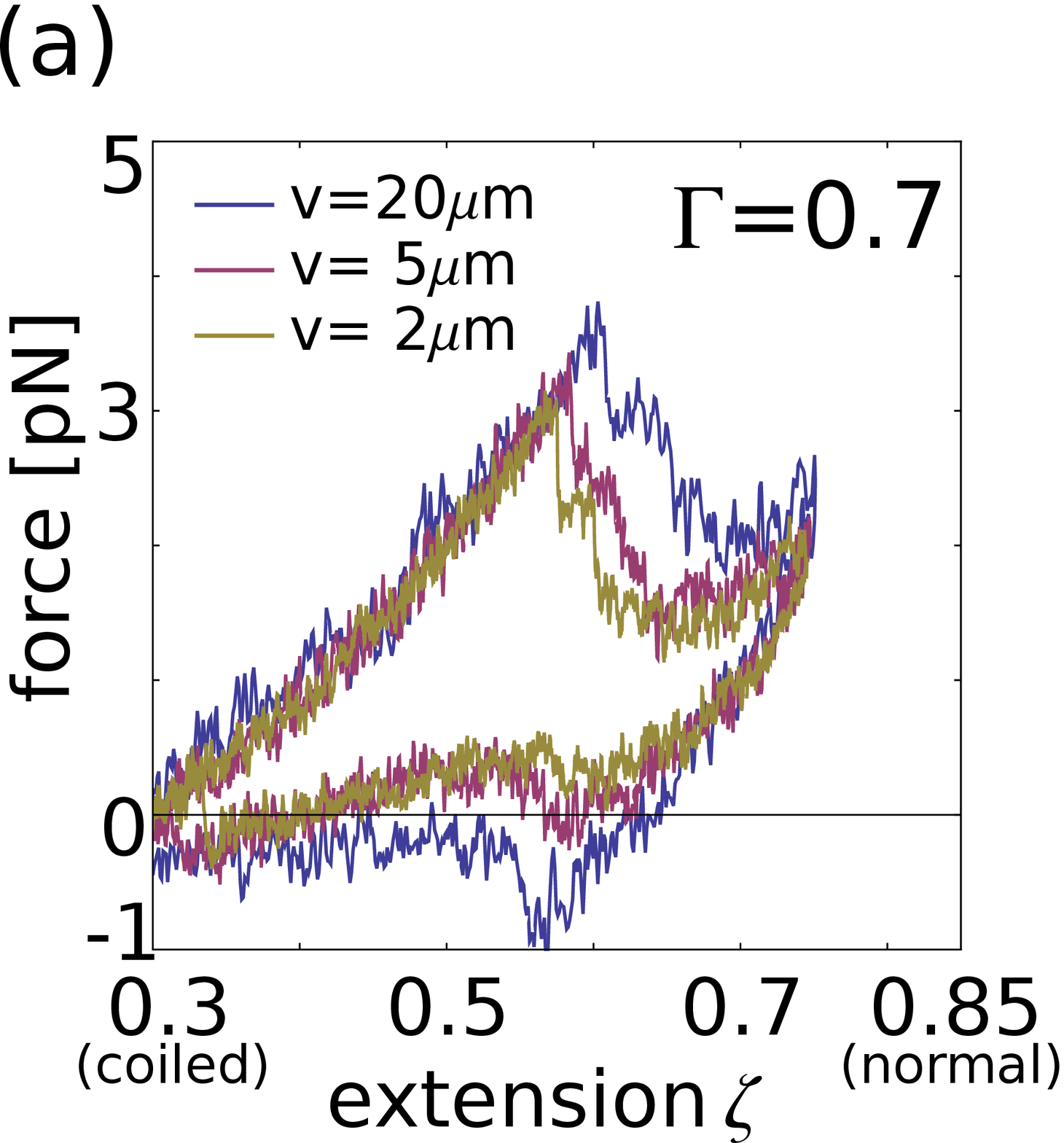}
	\includegraphics[width=.49 \columnwidth]{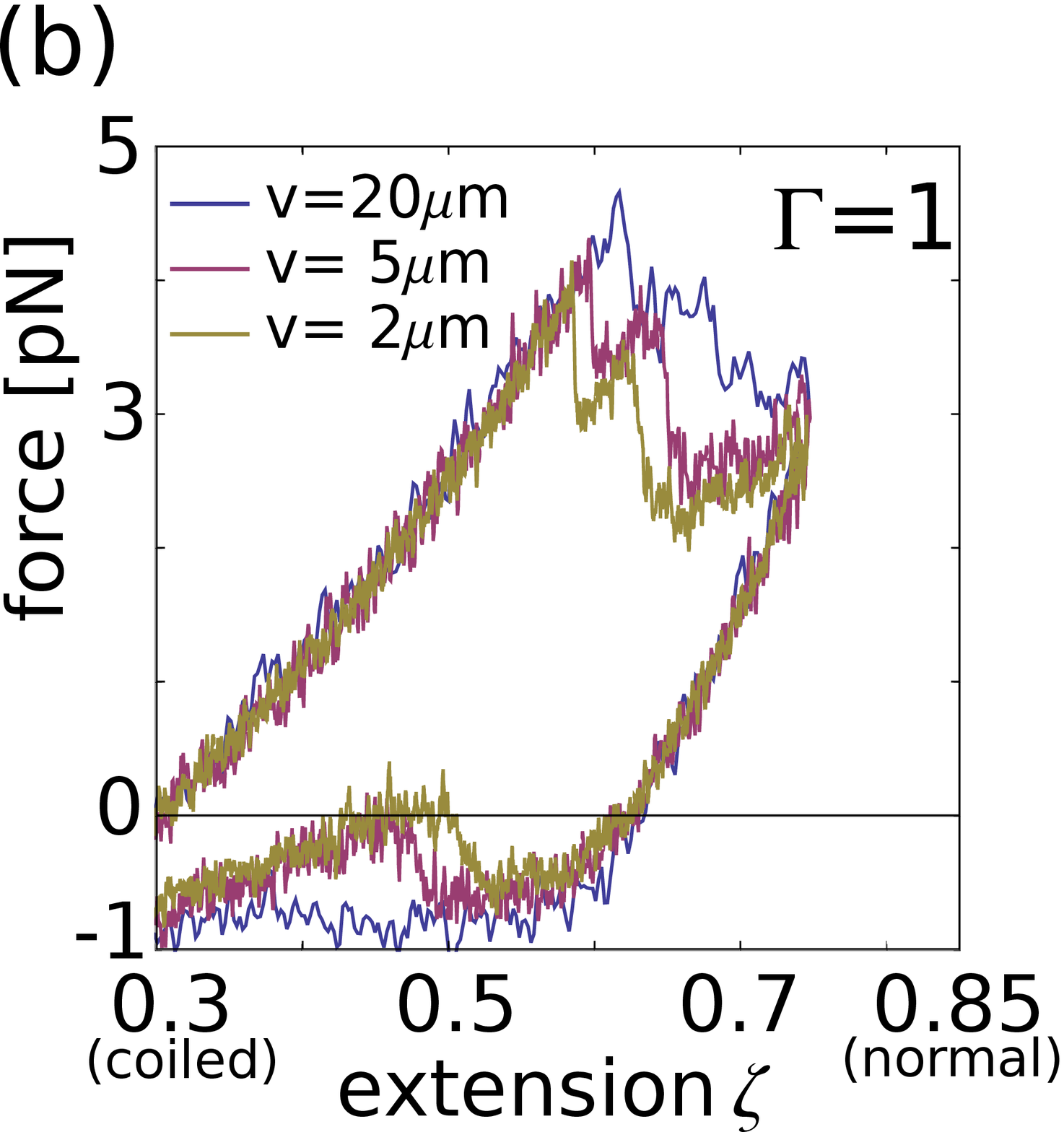}
    \end{center}
  \caption{Force-extension curves for twist-to-bend ratios $\Gamma=0.7$ (a)
and $\Gamma = 1$ (b) at different extension rates 
$v_p=2\micro\meter\per\second$, $5\micro\meter\per\second$ and 
$20\micro\meter\per\second$.}
  \label{fig: Velocity}
\end{figure}

In Section\ \ref{subsec: TimeScale} we reasoned that during 
the measurements of the force-extension curves the bacterial flagellum 
goes through a sequence of equilibrium states. This means that frictional 
forces acting on the filament through the solvent are negligible against 
elastic forces. Therefore, without thermal noise, the force-extension 
curve does not depend on the extensional rate $v_p$. In contrast, our
Brownian dynamics simulations demonstrate a clear influence of $v_p$. 
In Fig.\ \ref{fig: Velocity}, we show force-extension curves for 
twist-to-bend ratios $\Gamma=0.7$ and $1$ and different extension rates 
$v_p=2\micro\meter\per\second$, $5\micro\meter\per\second$ and 
$20\micro\meter\per\second$. The first transition from the coiled to
normal state occurs at larger extensions when $v_p$ is increased.
This is immediately clear since smaller velocities $v_p$ give the
filament more time to explore the energy landscape with the help of
thermal fluctuations. So the appropriate local curvature and torsion
to overcome the potential barrier can be created at smaller extension.
The curves in Fig.\ \ref{fig: Velocity} just give one specific realization 
for each parameter set. In the following section, we will investigate in
detail the probability distribution for the extension where the first
coiled-to-normal transition occurs. 

%

If the filament in the normal state is compressed too fast, it will start 
to buckle since again it has not sufficient time to overcome the energy
barrier.
This is visible in Fig. \ref{fig: Velocity}(a), where the
filament with the highest compressing rate $v_p = 20 \mu \mathrm{m/s}$
buckles which corresponds to a negative force. On the other hand,
for smaller rates $v_p$, the filament always returns into the coiled state.

%

\subsection{Clamped filament} \label{subsec.clamped}


Since the bacterial flagellum goes through a sequence of equilibrium 
states during the measurements of the force-extension curves,
it should be possible to derive some characteristics of 
these curves by investigating a clamped filament which is hold at a 
fixed extension $\zeta=z/L$ in 
thermal equilibrium. In particular, we show here how one can infer
the mean extension $\langle \zeta \rangle (v_p)$ at which the filament 
in a force-extension measurement undergoes the first local transition to 
the normal state  

\begin{figure}
\begin{center}
\includegraphics[width= 0.9 \columnwidth]{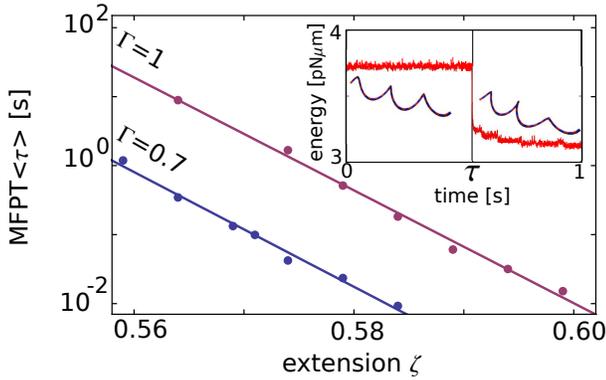}
\end{center}
\caption{Mean-first-passage time (MFPT) as a function of extension 
$\zeta$ for $\Gamma=0.7$ and $\Gamma=1.0$. Inset: When the local
coiled-to-normal transition occurs, the energy decreases instantaneously.}\label{fig: MFPT}
\end{figure}

We stretch the filament in the coiled state at zero temperature to an 
extension $\zeta$ and keep this extension constant. We then perform a 
Brownian dynamics simulation and determine for each realization the 
first-passage time $\tau$ at which a local transition to the normal
state occurs. The inset of Fig.\ \ref{fig: MFPT} shows the filament before and
after the transition accompanied by a sharp decrease of the elastic free
energy. According to Kramers theory, the mean-first-passage time 
(MFPT) is proportional to the Arrhenius factor,
\begin{align}
\m \tau \sim & \exp\left(\frac{\triangle \mathcal F(\zeta)}{k_B T}\right).
\end{align}
The activation energy $\triangle \mathcal F(\zeta)$ depends on the
extension $\zeta$ and is needed to create the domain wall between the
coiled and normal state. We determined the MFPT by averaging over
100 simulated values for $\tau$ at each extension $\zeta$.
The results are plotted in Fig.\ \ref{fig: MFPT} for two twist-to-bend
ratios $\Gamma=0.7$ and $\Gamma=1$. We only simulated $\m \tau$ for a small
variation in the extension for two reasons: at larger extensions where
$\triangle \mathcal F(\zeta) \approx k_B T$ the Kramers theory is no 
longer valid and $\m \tau$ is too small to be determined accurately;
at smaller extensions $\m \tau$ is so large that it cannot be calculated 
in reasonable simulation times. Note that for $\Gamma=1$ the MFPT spans 
three decades. Our simulation results demonstrate that 
$\log \m \tau (\zeta)$ can be fitted by
\begin{align}
 \log \m \tau (\zeta) \approx \alpha - \beta \zeta,
\label{eq.tau_m}
\end{align}
where $\m \tau$ is measured in seconds. Equation\ (\ref{eq.tau_m}) can 
just be viewed as a Taylor expansion to linear order in $\zeta$.
The parameters $\alpha$ and $\beta$ follow from a least-square fit.
Whereas $\alpha$ changes with $\Gamma$, the slope $\beta$ surprisingly 
does not seem to depend on $\Gamma$ within the numerical accuracy. 
A similar law as Eq.\ \eqref{eq.tau_m} is used in situations where bonds 
rupture under a given load force \cite{Evans2001,Bell1978}. Note, however, 
that here we control the extension.

\begin{figure}
\begin{center}
\includegraphics[width=.8 \columnwidth]{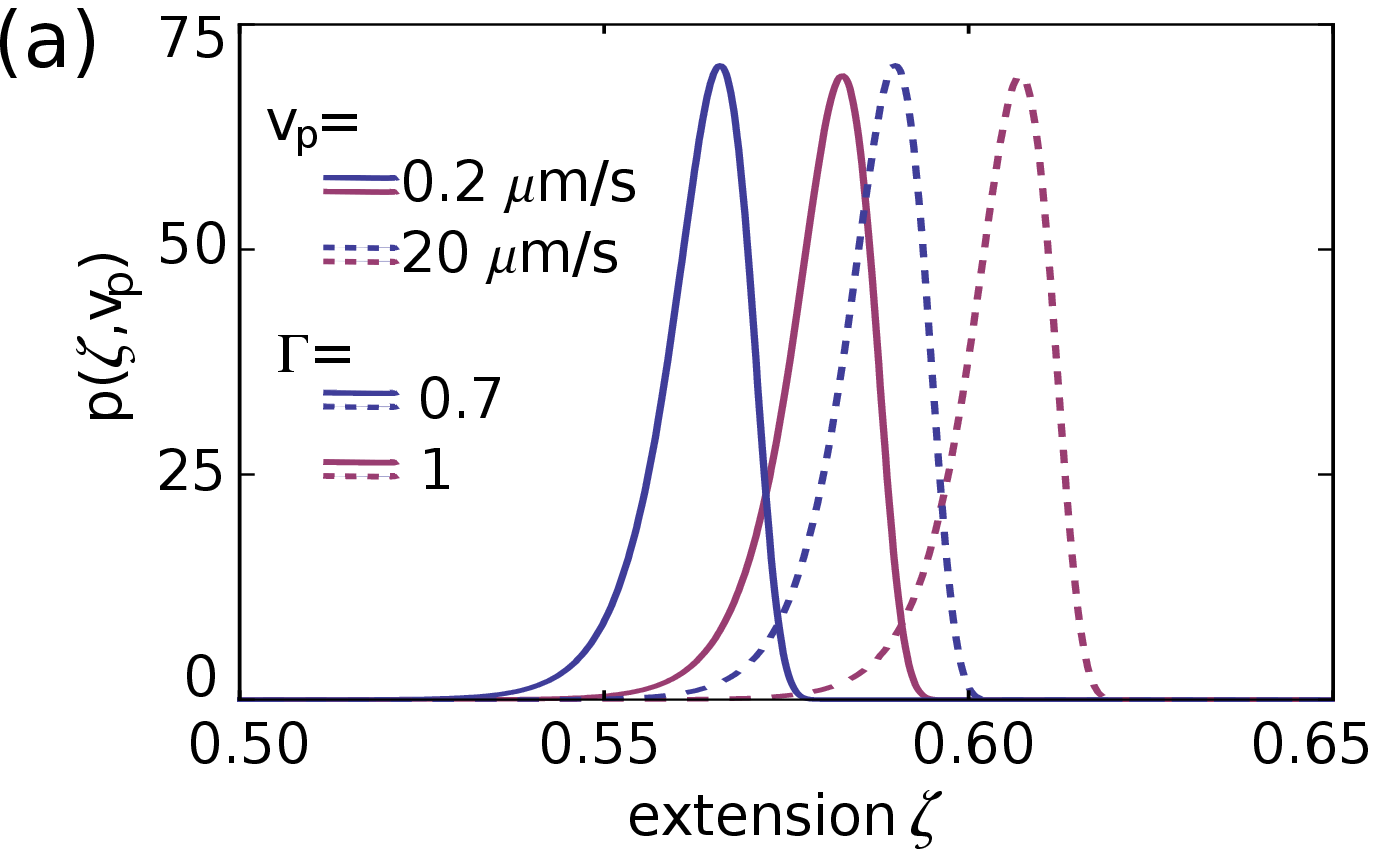}
\includegraphics[width=.8 \columnwidth]{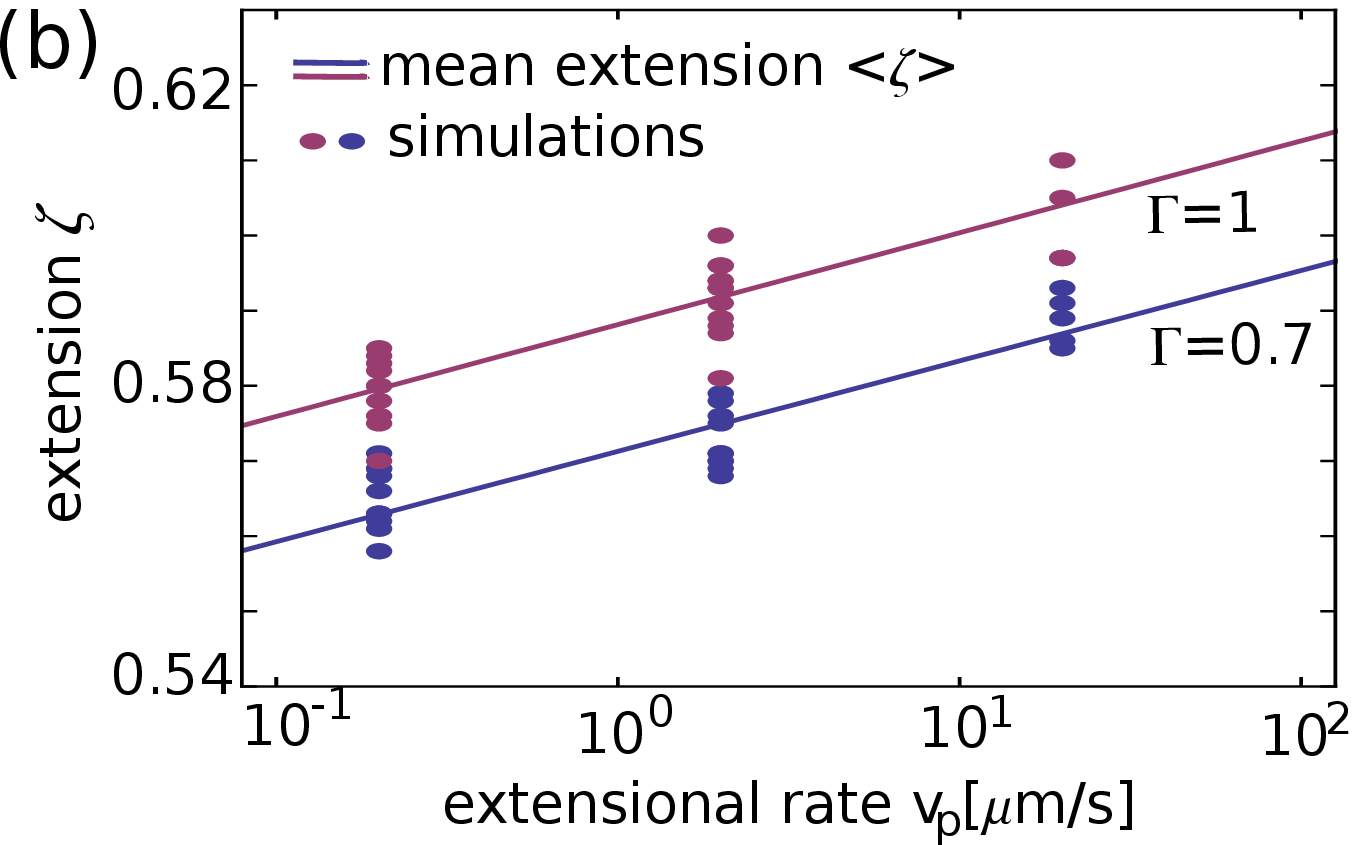}
\end{center}
\caption{(a) Probability distributions $p(\zeta,v_p)$ as a function 
of extension $\zeta$ for $\Gamma=0.7$ (blue) and $\Gamma=1$ (magenta)
at extension rates $v_p=0.2\micro\meter\per\second$ (full lines) and 
$v_p=20\micro\meter\per\second$ (dashed lines).
(b) Analytically determined mean extension $\m \zeta(v_p)$ for the 
coiled-to-normal transition as a function of velocity $v_p$ for 
$\Gamma=0.7$ (blue) and $\Gamma=1$ (magenta). The dots indicate 
numerically determined extensions for each simulation run.
}
\label{fig:meanext}
\end{figure}

Within the adiabatic or quasistationary approximation, we now formulate
the probability $p(t,v_p) \d t$ that with\-in the time interval $[t,t+\d t]$
the filament transforms locally from the coiled to normal state when 
stretched with velocity $v_p$,
\begin{align}
 p(t,v_p)&=\frac{1}{\m \tau(\zeta(t,v_p))}\exp\left(-\int_0^t 1/\m 
\tau(\zeta(t',v_p)) \d t'\right),
\label{eq.pt}
\end{align}
where 
\begin{equation}
\zeta(t,v_p)= \zeta_0 + t v_p/L
\label{eq.zetav}
\end{equation}
is the extension at time $t$.
The exponential factor in Eq.\ (\ref{eq.pt})
is the probability that until time $t$ a 
transition does not occur, which follows by writing the probability
that a transition within the time interval $\d t'$ does not happen as
$(1-\d t'/\m \tau) \approx \exp(-\d t'/\m \tau)$. The first factor is the 
probability per unit time that the transition occurs within $\d t$
at time $t$ when the filament is with certainty in the coiled state 
before. Using Eq.\ (\ref{eq.zetav}),
one introduces the probability $p(\zeta,v_p) \d\zeta = p(t,v_p) \d t$ that
the filament undergoes a local coiled-to-normal transition at 
extension $\zeta$. We calculate it analytically in Appendix\ 
\ref{app.transrate} by assuming the validity of Eq.\ (\ref{eq.tau_m}) 
for the whole $\zeta$ range. The results are plotted in 
Fig.\ \ref{fig:meanext}(a) for $\Gamma = 0.7$ (blue) and $\Gamma = 1$ (magenta)
and for extension rates $v_p=0.2\micro\meter\per\second$ (full lines) and 
$v_p=20\micro\meter\per\second$ (dashed lines). The probability 
distributions are concentrated on a small range about their maximum
values and are shifted to larger extensions for increasing velocities $v_p$,
as expected. We then calculate the mean extension 
\begin{equation}
\m \zeta(v_p) = \int_{\zeta_0}^{\infty} \zeta' p(\zeta',v_p) d \zeta',
\end{equation}
at which the coiled-normal transition occurs first for a given
extension rate $v_p$. The evaluation of the integral is performed 
in Appendix\ \ref{app.transrate} and finally gives
\begin{align}
\m \zeta(v_p) \propto \log v_p.
\label{eq.zetam}
\end{align}
The mean extension $\m \zeta(v_p)$ with all prefactors and constant 
terms is plotted in Fig.\ \ref{fig:meanext}(b) for $\Gamma = 0.7$ (blue) 
and $\Gamma = 1$ (magenta) together with numerically determined extensions
$\zeta$ for several realizations at a specific velocity $v_p$. The values 
scatter around the mean value and are, therefore, in good agreement with
the analytical treatment. We note that a relation similar to 
Eq.\ (\ref{eq.zetam}) occurs for the velocity dependence of the
rupture force of single molecular bonds in dynamic force spectroscopy\ 
\cite{Evans2001}.


\section{Summary and conclusions} \label{sec.sum}


\begin{figure*}[ht]
\begin{center}
\includegraphics[width= 1.7\columnwidth]{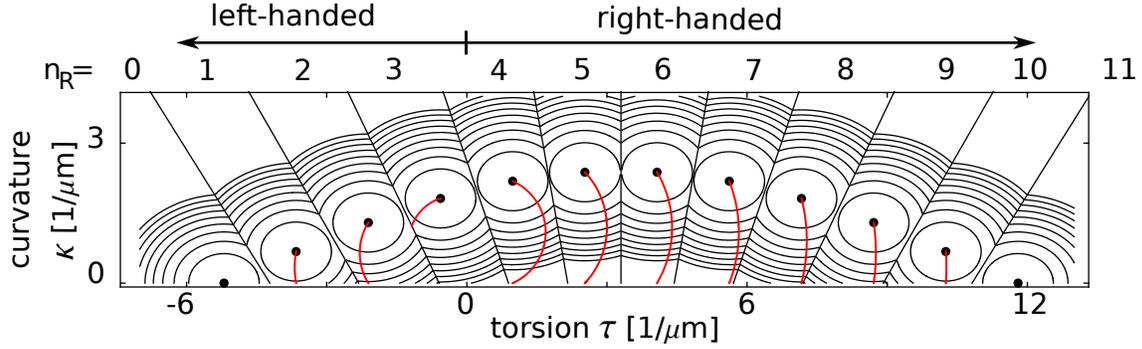}
\end{center}
\caption{Contour lines of the free energy density of the extended 
Kirchhoff rod theory that includes all 12 polymorphic states of
the bacterial flagellum. For all helical states the same ground state
energy and the same bending and torsional rigidity with $\Gamma=0.7$
is assumed. The red lines indicate the paths of uniformly stretched
flagella.}
\label{fig.extendedall}
\end{figure*}

In this article we have developed a sufficiently simple elastic model to 
describe the polymorphism of a bacterial flagellum based on Kirchhoff's 
theory of an elastic rod. The friction with the aqueous environment is 
modeled within resistive force theory. Using geometrical parameters of the 
coiled and normal states and the bending rigidity as obtained in 
Ref. \cite{Darnton07}, we are able to reproduce the force-extension curves 
recorded in experiments. Thermal fluctuations realized within Brownian 
dynamics simulations are crucial. In particular, the force values at which 
a first coiled-to-normal transition takes place lie between 3 and 5 pN, as in 
experiments \cite{Darnton07}. We have investigated in detail
how the force-extension curve depends on the twist-to-bend ratio $\Gamma$
and, furthermore, by analytic arguments identified a parameter region 
for ground-state energy difference $\delta$ and $\Gamma$, where a
coiled-to-normal transition should be observable. It clearly demonstrates 
that for values of $\Gamma$ well above one, a polymorphic transformation is 
not possible and therefore contradicts some of the experimental values for 
$\Gamma$ recorded in literature. Based on our simulations, we predict 
$\Gamma$ to be within $0.7$ and $1.0$. Further studies demonstrate how 
the extensional rate $v_p$ influences the force extension curve. We 
directly observe the influence in the simulations of the full 
force-extension cycle 
but also when we concentrate on the extension for the first coiled-to-normal 
transition. Since the extensional rates $v_p$ are sufficiently small, the 
flagellum goes through a sequence of quasi-stationary states. We, therefore, 
used equilibrium properties of clamped flagella to predict a logarithmic 
velocity dependence for the mean extension of the first coiled-to-normal 
transition in good agreement with our simulations.

Our approach is easily extended to include more than two polymorphic
states. Figure\ \ref{fig.extendedall} shows the contour lines of the 
resulting free energy landscape when all twelve polymorphic conformations 
of Fig.\ \ref{fig.polym} are taken into account. For each of these 
conformations 
we take Kirchhoff's elastic free energy with spontaneous curvature and 
torsion as given by Calladine\ \cite{Calladine75}
and just consider the minimum value from all 
these free energies. We assume here that the ground-state energies of 
all helical states are zero and that they all have the same bending and
torsional rigidities with $\Gamma = 0.7$. We also included the paths of
uniformly stretched helical filaments. The coiled-to-normal transition
($n_R=3$ to 2) takes place with certainty. 
However, all the other transitions would need 
thermal fluctuations to occur. For example, the normal-to-hyperextended
transition ($n_R=2$ to 1) will only occur when it is stretched sufficiently 
slowly so that thermal fluctuations can induce the transformation.

We can now use our model to study various aspects connected to bacterial 
locomotion. For example, we will investigate how polymorphic transitions are
induced by rotating the flagellum. Our very challenging goal is to model the 
complete tumbling cycle of a bacterium.
Finally, we note that our model may be applicable as well to spirochetes,
where very recent experiments and theory have begun to address the elasticity 
of the helical structure \cite{Dombrowski2009}.

\begin{acknowledgement}
We thank E. Frey, R. Netz, and A. Vilfan for stimulating discussions and 
acknowledge financial support from the VW foundation within the
program "Computational Soft Matter and Biophysics" (grant no. I/83 942).
\end{acknowledgement}


\appendix

\section{Extended Kirchhoff rod theory: the double-well potential} 
\label{sec.doublewell}

Here, the main idea is to extend the theory of Goldstein \textit{et al.}
\cite{Goldstein2000,Coombs2002}.
They used a double well potential for the twist density $\Omega_3$,
realized by a polynomial of degree four, to describe two polymorphic
states of a flagellum\ \cite{Goldstein2000,Coombs2002}.

In order to develop a strategy how to generalize this double well potential 
to all three coordinates $\Omega_i$ ($i=1,2,3$), we first write down a 
general one-dimensional polynomial of degree four:
\begin{align}
 f(x,x_1,x_2)&=\frac{A}{2}\frac{(x-x_1)^2}{(x_1-x_2)^2} [(x-x_2)^2 
\nonumber\\
&\qquad-\frac{d}{6} (x-x_1)(3x+x_1-4x_2) ]
\label{eq.poly}.
\end{align}
For $d<1$ it has two minima at $x_1$ and $x_2$ with $f(x_1)=0$ and 
$f(x_2)=\delta$, respectively, where 
\begin{align}
 \delta&=\frac{1}{12} A (x_1-x_2)^2 d.
\end{align}
Whereas $\partial_x^2 f(x_1) = A$, the second derivative at $x_2$ 
depends on the parameters $x_1$, $x_2$, $d$, and $A$.

\begin{figure}[t]
\begin{center}
	\includegraphics[width=1.0 \columnwidth]{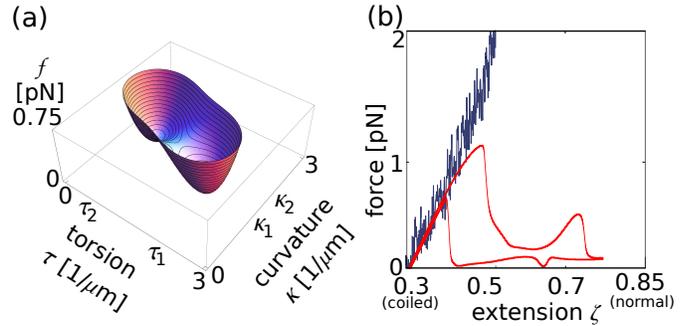}
\end{center}
\caption{(a) Double-well potential of Eq.\ (\ref{eq.poly3d}) plotted
as a function of $\Omega_2=\kappa$ and $\Omega_3=\tau$ for the parameters 
of the coiled and normal helical state. (b) Force-extension curve (red line)
simulated with this potential at $T=0$ and $v_p= 2\micro\meter\per\second$.
The blue curve is taken from Fig.\ \ref{Fig_A/C_Vergleich}(a) for the
same parameter $\Gamma=1$.}
\label{fig.doppelmulde}
\end{figure}

We generalize the polynomial of Eq.\ (\ref{eq.poly}) to three dimensions
by replacing terms of the form $Axy$ by $\vec x \cdot \mat A \vec y$,
where $\vec x$, $\vec y$ are three-dimensional vectors and $\mat A$
is a diagonal matrix with $A_{11}=A_{22}=A$ and $A_{33}=C$. The constants
$A$, $C$ are the bending and torsional rigidities, respectively.
Using the shorthand notation $|\vec x|^2_{\mat A} = \vec x \cdot \mat A \vec x$
with $\vec x = \vec \Omega - \vec \Omega_i$, we now write
\begin{align}
f(\vec \Omega,\vec \Omega_1,\vec \Omega_2)&
  =\frac{1}{2}\frac{|\vec \Omega-\vec \Omega_1|_{\mat A}^2}{ |\vec 
\Omega_1-\vec \Omega_2|_{\mat A}^2} [|\vec \Omega-\vec \Omega_2|_{\mat A}^2
\nonumber
\\&
\quad-\frac{d}{6}(\vec \Omega-\vec \Omega_1)\cdot\mat A(3\vec \Omega+\vec 
\Omega_1-4\vec \Omega_2) ] 
\label{eq.poly3d}\\
\text{with}\enspace \delta & =\frac{1}{12} 
|\vec \Omega_1-\vec \Omega_2|_{\mat A}^2 d.
\end{align}
The polynomial is illustrated in Fig.\ \ref{fig.doppelmulde}(a) with
$\vec \Omega = (0,\kappa,\tau)$. It has two minima at 
$\vec \Omega_1$, $\vec \Omega_2$ with
$f(\vec \Omega_1) =0$, $f(\vec \Omega_2) = \delta$, respectively, 
and one saddle point at 
$\vec \Omega_3 = \vec \Omega_1 + \frac{1}{2-d}(\vec \Omega_2-\vec \Omega_1)$ 
for $d<1$. Close to the first minimum at $\vec \Omega_1$,
the polynomial agrees with Kirchhoff's elastic free 
energy\ \eqref{Glg_freieEnergie2}, whereas the bending and torsional 
rigidities at $\vec \Omega_2$ and also the energy barrier depend on 
$\delta$.


Figure\ \ref{fig.doppelmulde}(b) shows a force-extension curve (red line)
simulated with Eq.\ (\ref{eq.poly3d}) at $T=0$ for the same parameters as 
in Fig.\ \ref{Fig_A/C_Vergleich}(a) with $\Gamma=1$. For comparison
the initial part of the curve from Fig.\ \ref{Fig_A/C_Vergleich}(a) 
($\Gamma=1$) is included. The helical filament is much softer 
and the initial part of the force-extension relation has a negative 
curvature in contrast to experiments. We, therefore, decided to introduce
the alternative model of Eq. (\ref{Glg Min}).


\section{Simplifying the free energy of Wada and Netz} \label{App_Energy}

After integrating out the spin degree of freedom, Wada and Netz arrived
at the following elastic free energy density \cite{Wada2008}:
\begin{align}
\label{eq_wadafreeenergy_1}
 f(\vec \Omega,\vec \Omega_1,\vec \Omega_2)&
  = f_1 - f_2 
\end{align}
with
\begin{align}
f_1=& 
  \frac{A}{2} \Omega_1^2 
    +\frac{A}{2}\left(\Omega_2 - \frac{\kappa_1+\kappa_2}{2}\right)^2
      +\frac{C}{2}\left(\Omega_3 - \frac{\tau_1+\tau_2}{2}\right)^2 \\
%
f_2=& \frac{k_B T}{a} \ln\Big[ \cosh(\Lambda) +
  \sqrt{\sinh^2(\Lambda)+\e^{-4J/k_B T}} \Big] + \frac{J}{a}
\end{align}
and
\begin{align}
\frac{k_B T}{a} \Lambda=&
  \delta 
+\frac{A}{2}(\kappa_1-\kappa_2)\left(\Omega_2-\frac{\kappa_1+\kappa_2}{2}
\right)\nonumber\\
&\qquad\qquad
      +\frac{C}{2}(\tau_1-\tau_2)\left(\Omega_3 - \frac{\tau_1+\tau_2} {2}\right)
\end{align}
Here $A,C$ are the bending and torsional rigidities, respectively, $\delta$ is 
the difference of the ground-state energies of the helical conformations
$(\kappa_1,\tau_1)$ and $(\kappa_2,\tau_2)$, and $a$ the length of
discretization. The quantity $J$ is the interaction strength in the 
Ising Hamiltonian. 

We interpret the energy cost $2J$ for two anti-parallel spins as the 
energy of a domain wall connecting two helical states. Since such 
domain walls are rarely seen in experiments \cite{Hasegawa82}, 
we can assume $2J \gg k_BT$ and, therefore, approximate $f_2$ as
\begin{align}
 f_2 & \approx \frac{k_B T}{a}|\Lambda|+ \frac{J}{a},
\end{align}
where we have used $\cosh x + |\sinh x|=\exp|x|$. Now, we introduce
the elastic free energy densities of the two helical states,
\begin{align}
\alpha &=\frac{A}{2} \Omega_1^2 +\frac{A}{2}\left(\Omega_2 - \kappa_2\right)^2 
+\frac{C}{2}\left(\Omega_3 - \tau_2\right)^2 +\delta/2,\\
\beta &=\frac{A}{2} \Omega_1^2 +\frac{A}{2}\left(\Omega_2 - \kappa_1\right)^2 
+\frac{C}{2}\left(\Omega_3 - \tau_1\right)^2 -\delta/2,
\end{align}
and write the two contributions to the energy density 
(\ref{eq_wadafreeenergy_1}) as 
\begin{align}
 f_1=&\frac{1}{2}(\alpha +\beta) + c_1, &
f_2\approx& \frac{1}{2}|\alpha-\beta|+ \frac{J}{a} ,
\end{align}
where $c_1$ is just a constant. This finally gives our ansatz for the
free energy density \eqref{Glg Min}:
\begin{align}
  f \approx &\frac{1}{2}(\alpha + \beta -|\alpha-\beta|)+ c_1 
 = \min(\alpha,\beta)+ c_1.
\end{align}

\section{Friction coefficients per unit length} \label{sec.friction}

In a moving helical filament, different parts interact via hydrodynamic
interactions. Nevertheless, using slender-body theory, Lighthill 
demonstrated that one can describe the hydrodynamic friction of the
filament with the help of resistive force theory that introduces local
friction coefficients per unit length. For the translational
coefficients, Lighthill obtained \cite{Lighthill1976}
\begin{align}
 \gamma_\parallel&= \frac{2 \pi \eta}{\ln(2q/r)} \enspace \mathrm{and} 
\enspace \gamma_\perp = \frac{4 \pi \eta}{\ln(2q/r)+1/2}.
\end{align}
Here $\eta$ is the shear viscosity, $r=0.02\micro\meter$ the cross-sectional 
radius of the bacterial flagellum, and $q$ a characteristic length, for which 
Lighthill derived $q=0.09\Lambda$, where 
$\Lambda=2 \pi/\sqrt{\kappa^2+\tau^2}$ is the filament length of one 
helical turn.
This gives $\gamma_\parallel\approx 1.54\eta$, $\gamma_\perp\approx 2.74\eta$
for the normal state and $\gamma_\parallel \approx 1.64\eta$, 
$\gamma_\perp\approx 2.91\eta$ for the coiled state. Since the coefficients
are similar in both states, we use the intermediate values 
$\gamma_\parallel\approx1.6\eta$ and $\gamma_\perp\approx2.8\eta$ for 
simulating the force-extension curves.
Finally, for the rotational friction coefficient one finds \cite{Brennen1977}
\begin{align}
\gamma_R &= 4 \pi\mu a^2.
\end{align}

\section{Velocity dependence of the mean extension $\m \zeta(v_p)$} 
\label{app.transrate}

In the following we use the numerically determined MFPT of 
Eq. (\ref{eq.tau_m}) in the form
\begin{equation}
 \m\tau = \tau_0 \exp[-\beta (\zeta-\zeta_0)].
\end{equation}
We transform the probability distribution $p(t)$ in Eq. (\ref{eq.pt})
into a probability distribution for the rescaled extension variable 
$x=\beta(\zeta-\zeta_0)= \beta v_p t/L $ using $p(t)\d t=p(x) \d x$.
The integral in Eq. (\ref{eq.pt}) is readily calculated with the help of
$\m\tau = \tau_0 \exp(-x)$ and we obtain
\begin{equation}
 p(x)\d x =\e^{x-A} \exp\left(-\e^{-A}(\e^x-1)\right)\d x,
\label{eq.px}
\end{equation}
where $\exp(-A)= L/(\beta v_p \tau_0)$. At $x=0$ or $\zeta=\zeta_0$,
the choice of our parameters shows that the energy barrier between 
coiled and normal state is much larger than $k_BT$, so $p(x=0)=e^{-A}$ 
is almost zero or $A\gg1$. Even for $x \approx 0$, we can therefore 
approximate the probability density in Eq. (\ref{eq.px}) as
\begin{equation}
p(x) \approx \e^{x-A} \exp\left(-\e^{x-A}\right)\d x:= p_0(x-A).
\label{eq.px2}
\end{equation}
It explains why all the probability densities $p(\zeta,v_P)$ in 
Fig.\ \ref{fig:meanext} have the same shape but are shifted relative 
to each other along the $\zeta$ axis due to different values of 
$v_p$ and $\Gamma$ and, therefore, of $A$. The distribution $p(x)$ has 
a maximum at $x=A$ with $p(A)\approx e^{-1}$.

We now calculate the rescaled mean extension
\begin{align}
 \m x &= \int_0^\infty x p(x) \d x.
\end{align}
We rewrite the probability distribution of Eq.\ (\ref{eq.px2}) as 
$p(x)=\e^{x-A} \exp\left(-\e^{x-A}\right)= -\@_x \exp\left(-\e^{x-A}\right)$
and obtain after integrating by parts
\begin{align}
  \m x &=\exp(\e^{-A})\int_0^\infty\exp\left(-\e^{x-A}\right)\d x.
\end{align}
The substitution $y=-\exp(x-A)$ then introduces the exponential integral 
function:
\begin{align}
 \m x&= - \exp(\e^{-A})\int_{-\infty}^{-\e^{-A}}\frac{\e^y}{y}\d y ,
\end{align}
which for $\e^{-A} \ll 1$ is approximated by
\begin{align}
 \m x &\approx -(C + \log\e^{-A}) = -C +A ,
\end{align}
where $C\approx 0.577$ is Euler's constant.
Introducing the original extension variable $\zeta$ and 
$A = \log(v_p \tau_0/L) + \log(\beta)$, we obtain for the mean extension 
at which the coiled-to-normal transition first takes place:
\begin{align}
\beta(\m\zeta - \zeta_0) = -C + \log(v_p \tau_0/L) + \log(\beta) \propto
\log(v_p) .
\end{align}

%

\end{document}